# Bursting Drops

Varun Kulkarni, Suhas Tamvada, Yashasvi Venkata Lolla, Sushant Anand

The material contained in this submission is published in the following two papers. (preprint) PDFs of both are attached with this file.

1. **Bursting of Underwater Oil Drops**
   V Kulkarni, VY Lolla, S Tamvada, S Anand (2024)
   Physical Review Letters 133 (3), 034004
   https://journals.aps.org/prl/abstract/10.1103/PhysRevLett.133.034004
   Open Access Link on google scholar

2. **Modulating outcomes of oil drops bursting at a water–air interface**
   Varun Kulkarni, Suhas Tamvada, Yashasvi Venkata Lolla, Sushant Anand (2024)
   Applied Physics Letters, 125 (14), 141601.
   https://doi.org/10.1063/5.0216820
   Open Access Link on google scholar

For any further information or materials connected with this work please contact the corresponding authors of the papers, Varun Kulkarni(varun14kul@gmail.com) and/or Sushant Anand (sushant@uic.edu).

# Bursting of Underwater Oil Drops

Varun Kulkarni, Venkata Yashasvi Lolla, Suhas Tamvada, and Sushant Anand*
*Department of Mechanical and Industrial Engineering*
*University of Illinois, Chicago, IL 60607*

For decades, two main facets of underwater oil spills have been explored extensively - the rise of oil drops and resulting evolution of the oil slick at the air-water interface. We report on the bursting of rising oil drops at an air-liquid interface which precedes slick formation and reveal a counter-intuitive *bulge* reversal that releases a daughter oil droplet inside the bulk as opposed to upward-shooting jets observed in bursting air bubbles. By unraveling the underlying physics we show that daughter droplet size and bulk liquid properties are correlated and their formation can be suppressed by an increase in the bulk viscosity.

Oil-water interactions lie at the root of diverse beneficial applications, such as drug delivery [1], oil recovery [2], food emulsions [3], and material synthesis [4], but they are also a subject of immense environmental concern due to oil seepages and spills [5, 6]. While oil seepages in oceans may date back to ancient times [7], our excessive reliance on oil and its transportation has increased their prevalence, worsening their environmental and economic impact [8, 9]. Consequently, for decades researchers worldwide have focused on studying major aspects of the oil spill process: rising of oil droplets through water[10], oil drops gently falling onto a liquid bath [6, 11, 12] and oil film spreading on an air-water interface [6, 13]. Although the coalescence at a liquid/liquid interface [14, 15] has been investigated, surprisingly, how an oil drop rising within the ocean transitions to an oil slick (forming a three-phase contact line) after breaching the air-water interface remains unexplored. This scenario bears similarities to the extensively studied phenomenon of bubbles bursting at an air-water [16–18] and air-oil/water interface[19]. For instance, in both cases, buoyancy plays a pivotal role in draining the liquid around the air bubble or oil drop. In air bubbles, the thin liquid film retracts in air[17], while for oil drops, the film retracts in the air on one side and over the oil drop's surface on the other [20], introducing significant viscosity effects in the latter case. Upon film retraction, bursting air bubbles produce jets releasing droplets upwards into the air. However, it is unclear whether the same can occur when oil drops burst upon breaching an air-liquid interface.

In this letter, we study the bursting process of oil drops at an air-liquid interface. We show that an oil drop doesn't always directly transition to forming an oil slick but can also form daughter droplets within the bulk liquid through a fascinating bulge reversal. Such a *reversal* (oriented parallel to gravity) is contrary to vertically rising jets (oriented anti-parallel to gravity) generated by bursting air bubbles[17] or liquid cylinders in falling drops[11] impacting liquids. The daughter droplet so released, cascades, forming a smaller droplet each time until the smallest droplet bursts to merge completely with the overlaying oil-slick. Although similar cascades have been reported for *falling* drops [6, 12, 14, 21, 22] they are known to experience drop distortion which releases a daughter droplet, $O(100\ \mu\text{m})$ only upwards unlike *rising* drops which we show can exhibit both downward drop distortion in the case of drops and jetting upwards for air bubbles, producing $O(100\ \mu\text{m})$ daughter droplets (see Supplementary Material [23] Sec. S1 and S2 for a detailed literature review and details of prevalent mechanisms). Using experimental and theoretical analysis, we establish the conditions for bulge reversal and downward droplet formation, demonstrate its control by merely changing the bulk viscosity and conclude by determining the scaling for daughter droplet sizes.

To explore the dynamics of drop bursting, we constructed an experimental setup (see Fig.1A, C) in which air bubbles or parent ($p$) oil drops of radius $R_p$ of different sizes emerging from a nozzle gently impact the bulk($b$) fluid/air($a$) interface (< 1 cm/s). We used hexadecane, pentane, and silicone oil as drops and glycerol/water mixtures 0-80 % wt. varied in increments of 10 % wt. as bulk fluids which have higher density, $\rho_b$ (in kg/m$^3$) than the oil drops, ($\rho_p$ in kg/m$^3$) with a density ratio, $\rho_r = \rho_b/\rho_p > 1$. The dynamic viscosity, $\mu_p$ was in the range of 0.24 to 3.005 mPas while $\mu_b$ varied between 1 to 60 mPas. Higher oil (drop) viscosity would dampen the capillary waves at their incipience and hence completely eliminate daughter droplet generation, therefore, in this work we consider any wave-damping only through a change in bulk viscosity. For the oils, spreading coefficient, $S = \sigma_{pa} - \sigma_{ba} - \sigma_{pb}$ (in N/m) with respect to water [24] where, $\sigma_{pb}$ represents the interfacial tension between the parent oil drop and bulk liquid, $\sigma_{pa}$ is the surface tension of oil and $\sigma_{ba}$ is the surface tension of bulk liquid, all expressed in N/m. $S > 0$ corresponds to oil spreading on the bulk forming a film (e.g. silicone oil and pentane), as opposed to $S < 0$, where the oil is non-spreading, taking a lenticular shape on the bulk (e.g. hexadecane) [6]. The oils were chosen such that they have similar physical properties [25, 26] to light crude oils and petroleum hydrocarbons found in oil spills. We combine these quantities to concisely describe our results in terms of two dimensionless groups, namely the Ohnesorge number, $Oh_b = \mu_b/\sqrt{\rho_b \sigma_{pb} R_p}$ of the bulk fluid and viscosity ratio of the bulk fluid to drop, $\mu_r = \mu_b/\mu_p$. Values for fluid properties, dimensionless groups and setup details are presented in Supplementary Material [23] Sec. S3. For visualization, the oil drops were stained using an oil-soluble dye - Sudan Blue II to easily

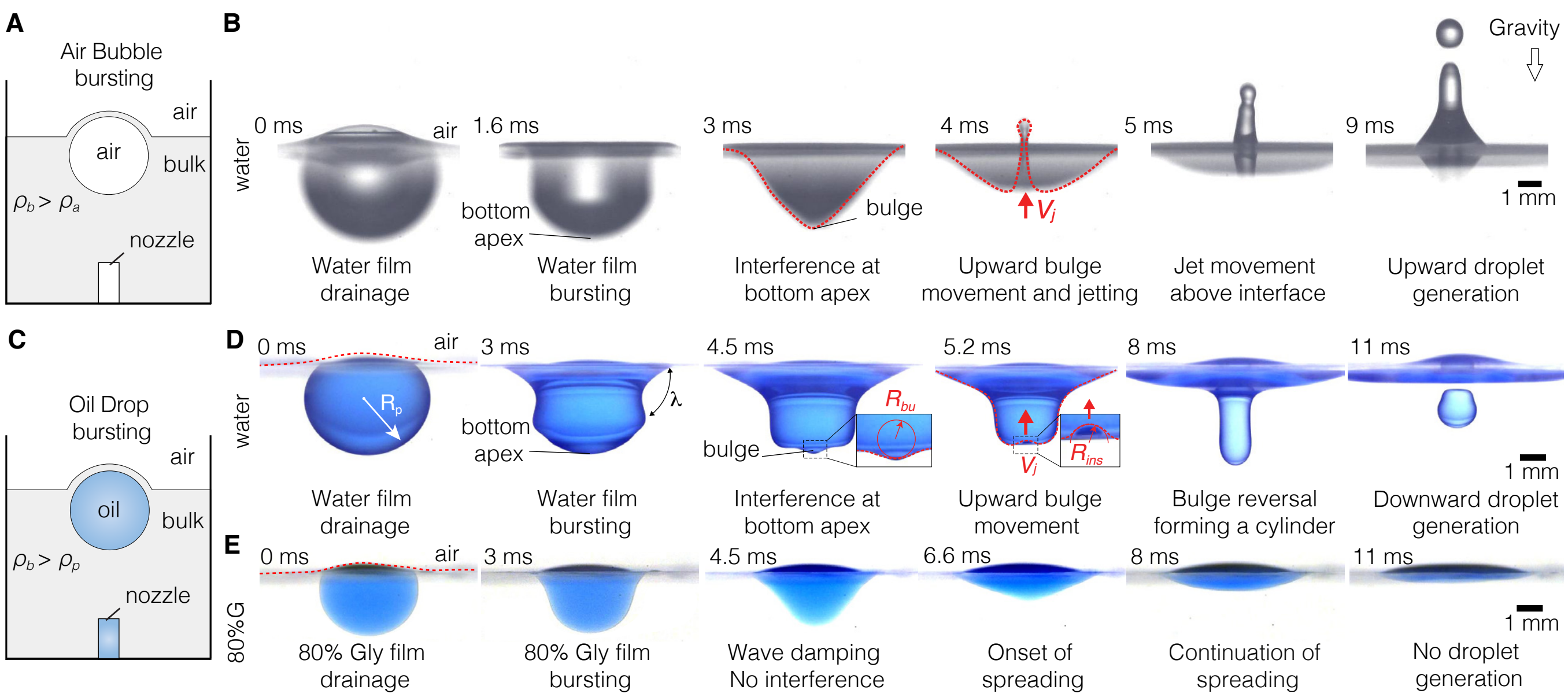

FIG. 1. **A.** Schematic depicting gently impacting air bubbles at a liquid-air interface **B.** Image sequence of a bursting air bubble showing *upward* jetting with bulge moving at velocity, $V_j$ at $t$ = 4 ms. **C.** Schematic depicting gently impacting oil drops at a liquid-air interface. Image sequence of a Hexadecane drop after bursting at **D.** water-air interface leading to bulge reversal and generation of a downward daughter droplet ($t$ = 11 ms) with bulge moving at velocity, $V_j$ at $t$ = 5.2 ms and, **E.** 80 wt. % glycerol-air interface leading to complete emergence without bulge reversal and upward daughter droplet formation.

differentiate them from the bulk liquid which was not dyed for any case (oil drops or air bubbles).

Fig.1B shows evolution of an air bubble bursting at an air-water interface. Following thin bulk liquid film rupture, capillary waves propagate downwards, leading to the formation of a highly curved region at its apex ($t$ = 3 ms). The momentum of this region produces an upward liquid jet ($t$ = 4 ms), which then pinches off to release daughter drops in the air ($t$ = 5 and 9 ms). In contrast, the temporal evolution of an oil drop bursting at the air-water interface is shown in Fig.1D. Here, on rupture of the liquid film, (see Fig.2A, **Video S1** and **S2**, Sec. S6) capillary waves rapidly descend straddling the oil-water interface while continuously being attenuated by viscosity of the drop and the bulk (see **Video S1**, Sec. S6). Upon reaching the bottom apex of the drop they constructively interfere to form a *bulge* (*bu*) [17] ($t$ = 4.5 ms) of radius of curvature, $R_{bu}$.

The *bulge* inverts due to an interfacial stress, $\sim \sigma_{pb}/R_{bu}$ forming a protrusion inside (*ins*) of a lower radius of curvature, $R_{ins}$ (at $t$ = 5.2 ms) and higher interfacial stress, $\sim \sigma_{pb}/R_{ins}$ before finally undergoing another reversal which pulls the drop downwards with it, parallel to gravity. The downward moving (deformed) drop takes a cylindrical liquid shape ($t$ = 8 ms) that ultimately pinches off, leaving behind a daughter droplet ($t$ = 11 ms) few 100 $\mu m$ in size, inside the water bath. The observation of *bulge reversal* in systems comprising oil drops rupturing at a air-water interface is independent of $S$ and has been unreported so far. To examine the controllability of this behavior, we substitute the water bath with glycerol/water solutions possessing higher viscosity in which capillary waves are strongly dissipated (typically > 50% by wt. glycerol) thereby precluding daughter droplet generation (Fig. 1E, **Video S1**,Sec. S6).

When bulge reversal produces a daughter droplet, a cascade of bursting events is triggered which produces smaller sized daughter droplets until complete emergence (not coalescence since drop and bulk phases are different) occurs with oil floating atop water (Fig.2B, also see **Video S3**, Sec. S6). From a practical standpoint, it indicates that drops bursting at an water-air interface from underwater oil spills can disperse daughter droplets deep inside the oceans. Furthermore, marine biosurfactants within the oceans [27, 28] can arrest daughter droplet generation cascade and prolong oil droplet lifetimes, aspects which can be explored in future studies. Finally, we are concerned with drop distortion which leads to daughter droplets of size, $O(100\mu\text{m})$ unlike previously reported [29] bubbles of secondary jets which are produced downward but much smaller in size $O(4-5\ \mu\text{m})$.

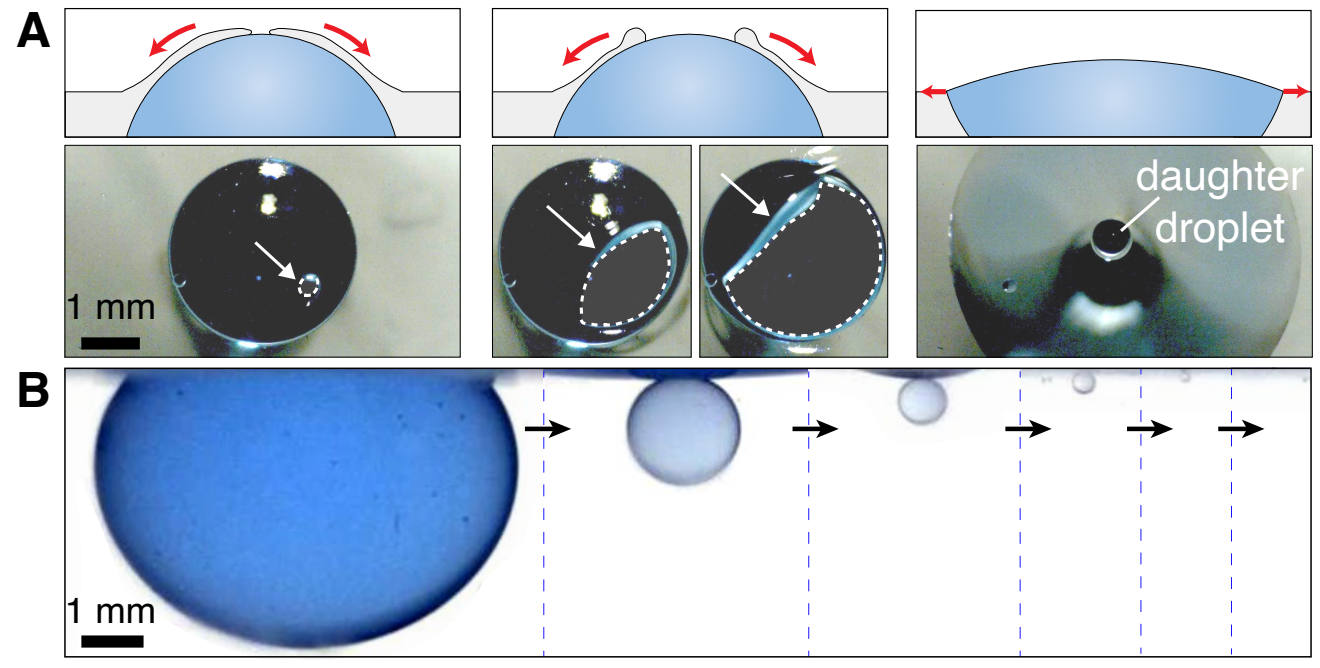

FIG. 2. **A.** Schematic (above) and top view (below) showing - hole formation, hole expansion, and spreading of drop with a submerged daughter droplet. **B.** Bursting cascade forming daughter droplets leading to complete emergence.

To understand in detail why the bulge reverses and distorts the drop downwards the we closely examine the bulge movement after it forms. So far we have broadly outlined interfacial tension stress as being important to bulge reversal. However, the associated interfacial energy opposes the kinetic energy of the bulge moving with a velocity $V_j$ immediately after its reversal (see Fig. 1D, 5.2 ms) similar to the description for bursting air bubbles[17] shown at a slightly later instant in Fig. 1B, 4 ms after the bulge has assumed the form of a discernible upward jet. $V_j$, strictly speaking, is a function of $\mu_b, \mu_p, \sigma_{pb}, R_p$ with the explicit dependence on these variables not considered for simplicity. The combined effect of these quantities [18, 31] on $V_j$ implies that for jets emanating from air bubbles its value is higher than for oil drops (see Supplementary Material [23] Sec. S2 for experimental values). This is because the bulge is slowed down by both viscosity of the oil drop and the bulk. Specifically, in the case of oil drops, a comparatively high interfacial tension stress develops at its tip (of radius $R_{ins}$) which carries a lower kinetic energy and forces it to reverse direction parallel to gravity, pinching off a daughter drop. Conversely, for an air bubble, the surface tension stress that develops at its tip (of radius $R_{ins}$) may be similar but possesses a higher kinetic energy which forces it to continue its movement vertically upwards [17, 30].

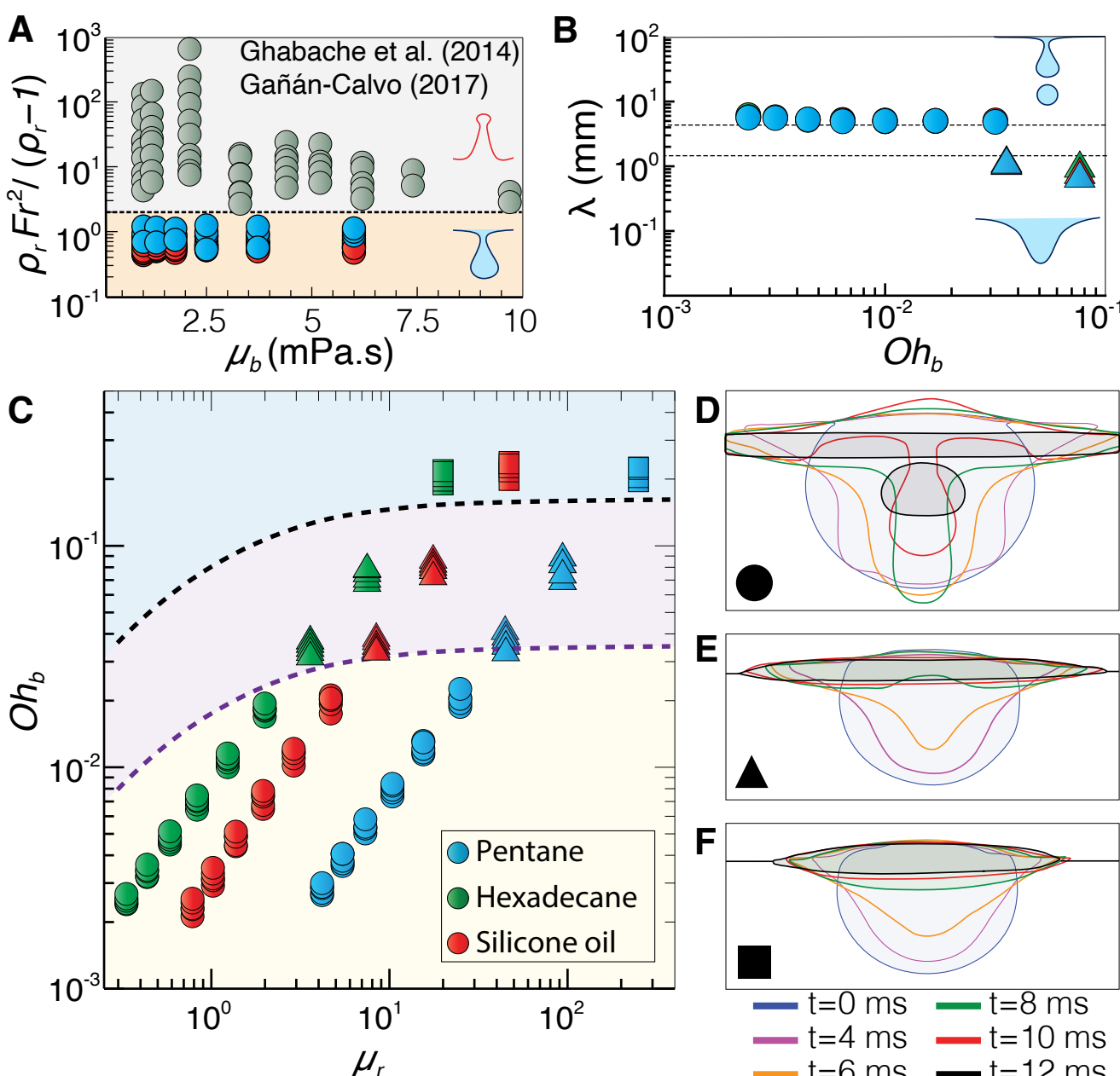

FIG. 3. **A.** Criterion for reverse (present study) and upward bulge movement [17, 30]. **B.** Wavelength of waves for bulge reversal with droplet generation or partial (filled circles) and delayed emergence (filled triangles).**C.** Different regimes: bulge reversal with daughter droplet generation, partial emergence (filled circles), bulge reversal with no daughter droplet generation, delayed emergence (filled triangles) and complete emergence (filled squares) as a function of $\mu_r$ and $Oh_b$. Drop profiles at different time instants for **D.** partial **E.** delayed and, **F.** complete emergence.

While the above is necessary for bulge reversal an overarching condition also needs to be met. This requires the excess kinetic energy (per unit volume) of the interfering waves, $\sim \rho_b V_j^2/2$ contained in the bulge when it moves upwards at velocity $V_j$ (see Fig.1B at $t = 4$ ms and Fig.1D at $t = 5.2$ ms) to be less than the gravitational energy (per unit volume), $(\rho_b - \rho_p) g R_p$ (accounting for buoyancy). With these considerations the condition for bulge reversal in dimensionless terms finally takes the form, $\rho_r(\rho_r-1)^{-1} Fr^2 < 2$ where, $Fr = V_j/\sqrt{gR_p}$ is the Froude number and $g$ is the acceleration due to gravity (shown in Fig.3A). The criterion developed here compares well with our experimental data and previous work[17, 30].

Having discussed the condition for *bulge reversal* we investigate how the strength of capillary waves that generate it, may or may not produce daughter droplets with *partial emergence* above the water-air interface of the remaining oil drop. To establish boundary of when this would happen we analyze the dynamics of capillary waves produced and their travel along the oil-bulk liquid interface. A typical wave is represented by its amplitude, $\xi_0$, complex frequency (with real, *re* and imaginary, *im* parts), $\omega (= \omega_{re} + i\omega_{im})$ and wave number, $k$ ($= 2\pi/\lambda$, where $\lambda$ is the wavelength) enabling depiction by the form $\xi = \xi_0 e^{\omega t + ikx}$ where $t$ and $x$ represent the time and spatial coordinate respectively.

When $\omega_{im} \neq 0$, we observe traveling waves that elongate the drop and characteristic of *bulge reversal*. To clearly identify when this leads to daughter droplet generation we henceforth use the term *partial emergence* and, when it does not, we use *delayed emergence* . For $\omega_{im} = 0$, monotonically decaying standing waves are observed without any drop elongation and the drop emerges above the air-bulk fluid interface completely, referred to as *complete emergence*. The real part of the complex frequency, $\omega_{re}$ is representative of the decay (or dissipation) rate of the waves. A dispersion relation for capillary waves which includes appropriate corrections for two superposed fluids [32] can be written in these terms as,

$$\omega = -Ak^2 \pm \sqrt{A^2k^4 - Bk^3 - Ck} \tag{1}$$

In the above, the first term, $-Ak^2$ represents $\omega_{re}$ or dissipation rate, $\mathcal{D}$ and the second term $\sqrt{A^2k^4 - Bk^3 - Ck}$ represents $\omega_{im}$ which relates to wave velocity as, $\upsilon = \omega_{im}/k$. Here, $A = 2(\mu_b/\rho_b)(1+\mu_r)(1+\rho_r) - 1$, $B = 2\sigma_{pb}/\rho_p(1+\rho_r) - 1$, $C = g(\rho_r - 1)/(\rho_r + 1)$. Detailed algebra leading to Eq. 1 appears in Supplementary Material [23], Sec. S4.

To demarcate *partial or delayed emergence* from *complete emergence* we inspect the second term inside the square root, $A^2k^4 - Bk^3 - Ck$ in Eq. 1. When this term is negative, $\omega_{im} \neq 0$, the waves travel with $\upsilon$ and we observe *partial* or *delayed emergence*. Conversely, if it is positive, the waves decay immediately leading to *complete emergence*. Here, the term $A^2k^4$ accounts for the slowing down of the waves due to viscosity [33] of oil drop and bulk fluid. Accordingly, the criterion for *partial or delayed emergence* can now be written as $A^2k^4 - Bk^3 - Ck < 0$. Gravity effect though important for bulge reversal does not influence the wavelength since the parent drop Bond number, $Bo_p = (\rho_b - \rho_p)R_p^2 g/\sigma_{pb} < 1$ (see Supplementary Material [23] Sec. S4) which implies $C = 0$ and $A < \sqrt{B/k}$. Substituting for $A$ and $B$ in terms of $Oh_b$, $\mu_r$, $\rho_r$, $kR_p$ and using the relation, $Oh_b = (\mu_r/\sqrt{\rho_r})Oh_p$ we obtain,

$$Oh_b < \sqrt{\frac{1+\rho_r}{4kR_p\rho_r}}\frac{\mu_r}{1+\mu_r} \tag{2}$$

Since $\rho_r \approx 1.55$ for our test conditions (see Supplementary Material [23] Sec. S2 and S4), the prefactor to $\mu_r/(1+\mu_r)$ in Eq. 2 reduces to $\sqrt{0.4/kR_p}$. To find $k$, we experimentally measure $\lambda$, the distance between two successive crests of the traveling wave train (see Fig. 1D, $t = 3$ ms, Supplementary Material [23] Sec. S4) near the base. We find $\lambda$ = constant = 1.34 mm across a range of $Oh_b$ near the transition boundary. $kR_p$ is therefore around 12.4 for $R_p \approx 2.65$ mm (see Fig.3B) giving a value 0.18 for the prefactor. This boundary is shown by black dotted (upper) curve in Fig. 3C.

Next, we establish the boundary between *partial* and *delayed emergence* (see Fig.3C, purple dotted (lower) curve) once Eq. 2 is satisfied. Prior studies [16, 17] on bubble bursting state that for bulge movement without reversal or, jetting to occur, the waves should not dissipate in amplitude completely before reaching the bottom apex *i.e.* $\mathcal{D}t_{base} < 1$ [12, 16, 21] where, $t_{base}$ is the time the waves take to reach the bottom apex. The length ($L$) the wave needs to travel to reach the base is $(\pi/2)R_p$ which can be used to calculate $t_{base}$ as $L/\upsilon$. We use the same for downward traveling waves due to their generality and from Eq. 1 deduce, $\mathcal{D} = Ak^2$ and $\omega_{im} = \sqrt{A^2k^4 + Bk^3 + Ck}$. After substituting for A, B, and C and neglecting gravity for small $Bo_p < 1$, we obtain the following criterion,

$$Oh_b < \sqrt{\frac{1+\rho_r}{4kR_p(\pi^2k^2R_p^2+1)\rho_r}}\frac{\mu_r}{1+\mu_r} \tag{3}$$

For the tested oils $\rho_r \approx 1.55$ and at transition to *partial emergence* we found the experimental value of $\lambda = 4.6$ mm measured in a similar manner (see Supplementary Material [23] Sec. S4) as before we get values of $kR_p \approx 3.7$ for $R_p \approx 2.65$ mm (see Fig.3B, triangles). This yields an inequality like the previous criterion but with a constant prefactor of 0.028 (for more details see Supplementary Material [23] Sec. S3 and S4) where the theoretical regime boundary is represented by the purple dashed line in Fig.3C. To validate Eq. 3, we compare it with the case of an air bubble bursting in water using, $\mu_r >> 1$, $\rho_r >> 1$, and $\pi k^2R_p^2 >> 1$. Since the wavelength is small due to the low viscosity of water Eq. 3 now simplifies to $2\pi Oh_b(kR_p)^{3/2} < 1$ recovering previous results exactly [12, 21].

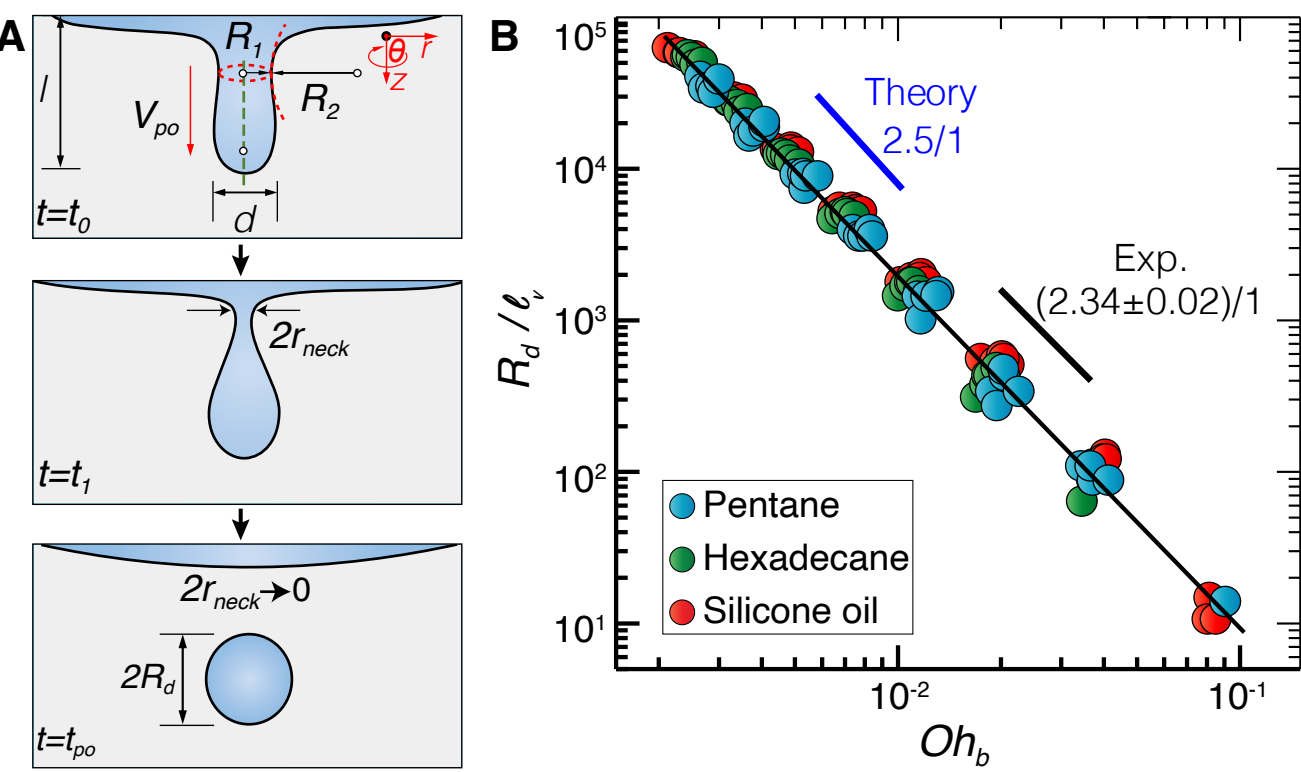

FIG. 4. **A.** Schematic showing drop profile at three time instances, $t_0$, when a cylindrical entity is formed, $t_1$, during thinning and, $t_{po}$ when a daughter droplet is pinched off. **B.** Scaling for the dimensionless daughter droplet radius, $R_d/\ell_v$ with $Oh_b$.

Time resolved experimental outlines for different regimes are shown in Fig.3D, E and F for the three oils tested. The profiles in Fig.3E (filled triangles) and Fig.3F (filled squares) do not produce droplets; however, we can see a relatively sharper traveling wave at $t = 6$ ms (orange line) Fig.3F in comparison to the one for the standing decaying wave in Fig. 3(F) which served to distinguish visually *partial or delayed* from *complete emergence*. Note that we did not observe satellite droplet formation and our results directly relate to oil spills in water bodies which often have varying bulk viscosities due to their composition, salinity, and temperature[34]. Further, given the broad nature of the derived criteria Eq. feqn2 and Eq. 3, we envision these to be applicable to coalescence at liquid-liquid interfaces too[35].

A salient aspect of *partial emergence* is generation of a daughter droplet of size, $R_d$ whose scaling we determine next. Fig.4A schematically shows the deformed drop just before and after the pinch-off. After the waves interfere at the base at $t = t_0$ a cylindrical entity of diameter, $d$, and length, $l$ is formed that moves downward with an average velocity, $V_{po} = l/t_{po}$ and thins continuously as shown at $t = t_1$, eventually pinching off in time $t = t_{po} = \sqrt{\rho_p R_p^3/\sigma_{pb}}$. The thinning initially is a result of downward movement of the cylinder and is accompanied by horizontal spreading above the interface such that the two motions compete against each other forming a neck with the final pinch off being capillary driven [11, 12]. The pressure difference inside and outside the neck, $p_{in} - p_{out} = \Delta p_{neck}$ during thinning equals $-r_{neck}^{-1}\sigma_{pb}$ as $r_{neck} \to 0$ and $R_1, R_2 >> r_{neck}$ pushing oil away (vertically) from the neck ultimately generating a daughter droplet of volume, $(4/3)\pi R_d^3$ which equals the volume of cylindrical entity, $\pi d^2 l/4$ before pinch-off.

The scaling law for $R_d$ hence can be ascertained once the scaling for $l$ and $d$ is known. To begin we assume the flow inside the cylindrical entity as near inviscid while that outside (bulk) to be viscous since $\mu_r > 1$.

At the moving cylindrical liquid front the viscous, $F_v \sim (\mu_b V_{po}/R_p)ld$ and capillary force $F_c \sim \sigma_{pb} d$ are balanced yielding, $l \sim (\sigma_{pb} R_p/\mu_b)V_{po}^{-1}$. Simplifying further we get, $l^2 \sim (\rho_r/Oh_b)^{1/2} R_p^2$. For, $\rho_r \approx 1.3$ and using the scale $\ell_v = \mu_b^2/\rho_b \sigma_{pb}$ [17] for making $l$ dimensionless, since the bulge leading to formation of daughter droplet is analogous to bubble bursting we eventually obtain, $l/\ell_v \sim Oh_b^{-5/2}$. As the energy of waves at the bottom apex decreases with an increase in bulk viscosity such that $l \sim d$, confirmed *a posteriori* by our experimental data (see Supplementary Material [23] Sec. S5) which show exponents of −2.23 and −2.25 for $l$ and $d$ respectively. The scaling for $R_d$ using the previously stated mass conservation results in, $R_d/\ell_v \sim Oh_b^{-5/2}$. The scaling exponent of −5/2 shown in Fig.4B is within 7% of −2.34 obtained from fitting. The close agreement between theory and experiments despite assuming inviscid flow inside the drop shows that viscous effects arising from the drop may be insignificant for our experimental conditions.

To summarize, this work shows that the bursting of a rising oil droplet at a liquid-air interface can lead to previously unreported bulge reversal and daughter droplet formation which are distinct from observations on coalescence of oil drops gently falling on an air-liquid interface. We delineate the different outcomes in a regime map based on $\mu_r$, and $Oh_b$ alone. Most studies heretofore have focused on oil-slick formation on the interface however droplet generation under it suggests a new pathway for oil spill proliferation which can adversely affect aquatic life. Moreover, presence of surfactants can arrest daughter droplet cascade ensuring that the oil droplets remain inside the bulk for long periods. In addition to guiding oil spill remediation our results could potentially be beneficial to numerous fields where oil-water interactions are important, these include oil recovery, drug delivery, food and cosmetics.

Financial support by National Science Foundation EAGER grant no. 2028571, UIC College of Engineering and Society in Science, Branco Weiss fellowship is gratefully acknowledged.

---

* Corresponding Author: sushant@uic.edu

# Modulating outcomes of oil drops bursting at a water-air interface

Varun Kulkarni,[a] Suhas Tamvada, Yashasvi Venkata Lolla, and Sushant Anand[b]
*Department of Mechanical and Industrial Engineering, University of Illinois at Chicago, Chicago, IL 60607, USA*

Recent studies have shown that capillary waves generated by bursting of an oil drop at the water-air interface produces a daughter droplet inside the bath while part of it floats above it. Successive bursting events produce next generations of daughter droplets, gradually diminishing in size until the entire volume of oil rests atop the water-air interface. In this work, we demonstrate two different ways to modulate this process by modifying the constitution of the drop. Firstly, we introduce hydrophilic clay particles inside the parent oil drop and show that it arrests the cascade of daughter droplet generation preventing it from floating over the water-air interface. Secondly, we show that bursting behavior can be modified by a compound water-oil-air interface made of a film of oil with finite thickness and design a regime map which displays each of these outcomes. We underpin both of these demonstrations by theoretical arguments providing criteria to predict outcomes resulting therein. Lastly, all our scenarios have a direct relation to control of oil-water separation and stability of emulsified solutions in a wide variety of applications which include drug delivery, enhanced oil recovery, oil spills and food processing where a dispersed oil phase tries to separate from a continuous phase.
Area: Interdisciplinary Applied Physics, Surfaces and Interfaces.

Dispersion of oil droplets in aqueous solutions are common in household items like salad dressings[1], ointments[2] and cosmetic creams[3] but even at larger scales in environment and industry such as oil spills[4,5] and enhanced oil recovery[6]. These interactions are characterized by either their ability to remain together as a homogeneous mixture or get separated as disparate oil and water phases. In this light, two major directions for research have emerged over the years, on the one hand, the focus has been on understanding the kinetics and enhancing stability of dispersed oil droplets in a continuous water phase[7–9] while on the other, mechanisms for their separation have been investigated. Results from these findings have often found use in applications like emulsion synthesis[10] and dispersal of oil slicks modulated by surfactants[11], microbes[12], particles[13] or the combination thereof[14]. While the role of chemical constitution and its symbiotic relationship with physical hydrodynamics in separation or homogenizing of oil-water mixtures using physio-chemical modifiers like surfactants or particles is known[15,16], it is yet unclear whether purely hydrodynamic mechanisms, which rely only on interfacial forces can be the main drivers of such two-phase interactions.

Along this line, recent work by *Kulkarni et al.*[17,18] has revealed a novel, unexplored pathway of oil separation from the surrounding water phase by investigating the bursting of a rising oil drop at an air-water interface. It was demonstrated that the bursting oil drop produces a daughter droplet within the continuous phase and the process cascades down until the oil drop form a film above the water-air interface. The significance of this may be readily appreciated, as leakages from broken underwater oil pipelines and natural seeps release oil droplet plumes where individual oil drops rise[19], eventually bursting at the water-air interface and may even be encountered in two-phase microfluidic droplet-based flows. Quite naturally, methods to control this separation provide interesting avenues for future investigations, with profound implications.

In this work, we pursue this objective and demonstrate two specific methods to control and modify this behavior by manipulating the oil drop/water interface which heretofore has only been shown by changing the viscosity of the water bath[17,18](continuous phase). In the first, we show that hydrophilic Bentonite (clay) particles mixed in an oil drop rising in a water bath can self-assemble at the oil-water interface to controllably halt successive bursting events. This situation resembles crude oil leaking from underwater pipeline bursts and entrains mineral/clay particles from the seafloor[20–22] or that seen in separation of Pickering emulsions[23]. Our second study is inspired by double emulsion synthesis and buoyant crude oil jets entrapping the surrounding water phase to form a water-in-oil compound drop[24]. We consider bursting of such compound drops made of encapsulated water drops of varying diameters which heretofore been only investigated for oil-coated air bubbles[25–27] or falling drops[28]. By these simple modifications to the oil drop-water interface we may regulate bursting outcomes such as cessation of bursting cascade at any given stage corresponding to a particular daughter droplet size and morphology of the daughter droplet produced.

For our experiments[17,18], we use hexadecane ($\rho_o$, 773 kg·m$^{-3}$ and dynamic viscosity, $\mu_o$, $3\times10^{-3}$ Pa·s) oil drops, insoluble in a D.I. (deionised) water bath ($\rho_w$, 998 kg·m$^{-3}$ and dynamic viscosity, $\mu_w$, $1\times10^{-3}$ Pa·s). In addition to its practical relevance, since water or hexadecane do not significantly dampen the capillary waves as in the case of more viscous outer liquids or typical two-fluid drop dynamics[29,30], we choose them as the continuous (bulk) liquid and dispersed (drop) phase, respectively to observe all subsurface phenomena clearly. The interfacial tension, $\sigma_{w/o}$ between hexadecane and water is 0.052 N·m$^{-1}$ and $\mu_o$ of hexadecane is small enough for viscous effects to be neglected[4]. Furthermore, hexadecane has a spreading coefficient of -0.0083 N·m$^{-1}$ which makes it non-spreading, forming a lenticular shape above the water-air interface. Its spreading is also considered negligible since its time scale is much longer than that for the subsurface dynamics[31]. The experimental observations are recorded using videos[4] taken at 4500 fps and a resolution of 1024 × 1024 pixels such that 1 pixel ≈ 15 $\mu m$. The depth of the water bath is maintained such that when the drops are completely detached from the needle they continue to be completely submerged inside the water bath. Additional details of the experiments are presented below in the appropriate sections.

---

[a] Corresponding author, electronic mail: varun14kul@gmail.com
[b] Corresponding author, electronic mail: sushant@uic.edu

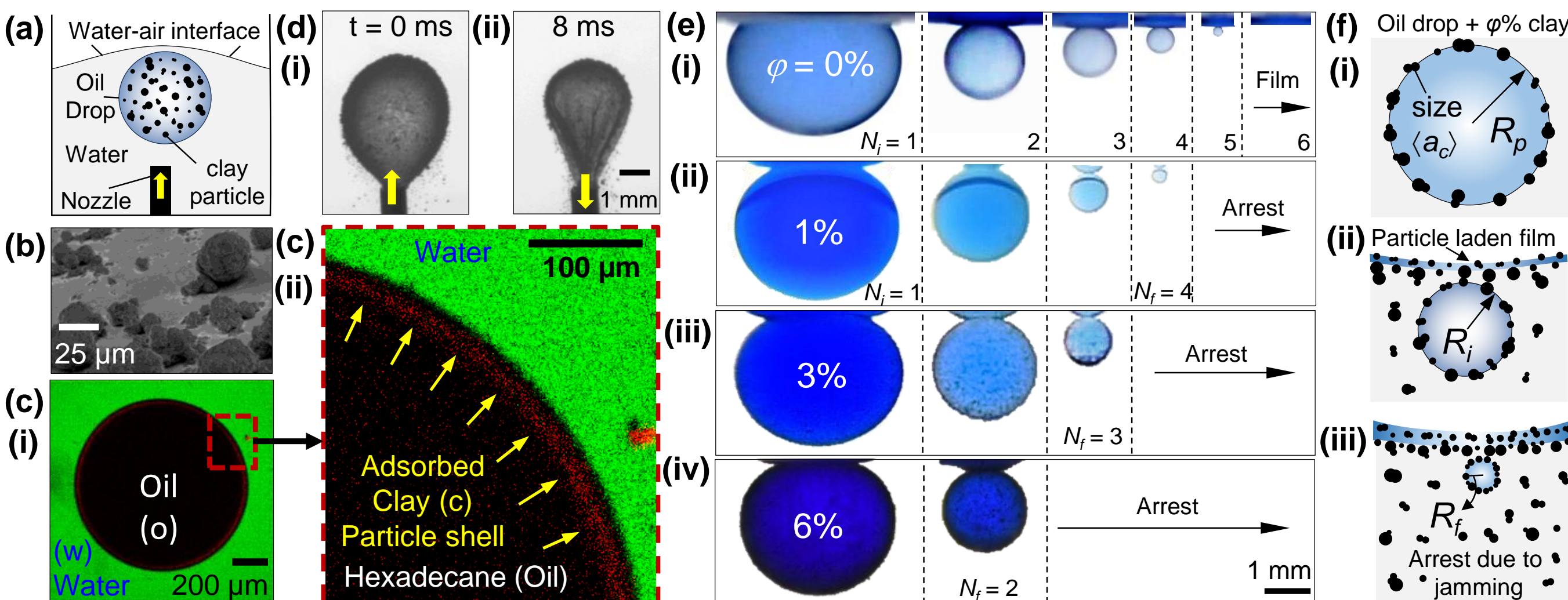

FIG. 1. (*a*) Schematic of experimental setup showing a rising oil drop containing bentonite clay particles bursting at an air-water interface after being released upwards. (*b*) Scanning electron microscopy (SEM) image of bentonite clay particles with a median size of 20 $\mu$m. (*c*)-(*i*) Confocal laser scanning microscope (CLSM) image of an hexadecane (oil) drop in water with adsorbed clay particles at the oil-water interface forming a shell. Clay particles and bulk water are labeled using Rhodamine-B fluorescent red dye and Fluorescein green dye respectively with hexadecane not being dyed (therefore seen in black). (*c*) -(*ii*) Zoomed view of (*c*)-(*i*) showing the adsorption of particles at the oil-water interface. Yellow arrows show the shell of clay particles (dyed red) (*d*) Proof of self assembly of clay particles demonstrated by, (*d*)-(*i*) injecting a particle-laden oil drop into a bulk of water and, retracting the liquid (*d*)-(*ii*). Self assembled particles adsorbed at the oil-water interface form a shell around the oil droplet which crumples during retraction. Effect on cascade(Multimedia available online), (*e*)(*i*) without particles[17], (*ii*)-(*iv*) with particles showing early cessation with increasing particle concentration, $\varphi$ = 1, 3 and 6 % (*f*) Mechanism of cascade arrest, (*i*) Initial particle coverage (*ii*) Drop bursting leading to particle-laden oil film and daughter droplet generation with increased surface coverage below the interface (*iii*) Formation of a *Pickering* drop with final arrest with 90% drop surface coverage of particles.

To investigate the effect of inclusion of particles on oil drop bursting at the water-air interface a known initial weight, $\varphi \approx$ 3% of hydrophilic bentonite clay particles was introduced into the parent ($p$) oil (hexadecane) drop as illustrated in Fig. 1($a$). The diameter of the clay ($c$) particles measured using SEM varied between 5−25 $\mu m$ with an average value, $\langle a_c \rangle$ of 15 $\mu m$, where, $\langle \cdot \rangle$ stands for the average of particles of different sizes (see Fig. 1($b$)). The particle-laden oil drop is released from a 0.5 mm inner diameter nozzle producing a parent drop of radius, $R_p$ equal to 2.2 mm and volume of nearly 45 $\mu l$. The bentonite clay particles are hydrophilic in nature[32] and almost completed wetted by water due to which the oil-water interfacial tension (and surface energy) is not affected much and we only use them in weight percentages between 0-6% which does not alter the overall oil drop density, significantly. Lastly, these particles have low solubility in water but their negative surface charge is neutralized by it, depressing their zeta potential[33] and not considered as a dominant factor in our analysis

As the oil drop rises through water due to the hydrophilic nature of the clay particles they are expected to adsorb[34] and self-assemble at the oil-water interface. To confirm this we conducted two specific experiments. In our first study, we used confocal laser scanning microscopy (CLSM) with water labeled using a green fluorophore and clay particles using a red one. The self-assembly of the clay particles upon release of particle-laden oil drop to form a shell (red ring) encapsulating the drop surrounded by water is clearly visualized here and shown in 1($c$)($i$) (taken mid-plane, cutting across the diameter of the drop). The enlarged view in 1($c$)($ii$) shows details of part of the drop revealing the formation of a 20 $\mu m$ thick particle shell covering the oil drop of radius 1 mm.

To further confirm these observations we conducted a second set of experiments in which we dispensed an oil drop containing clay particles from a nozzle allowing the hydrophilic particles to self-assemble at the oil-water interface and form a shell as depicted in Fig. 1($d$)($i$), $t = 0$ $ms$. Thereafter, we slowly retract oil from the nozzle and notice discernible creasing/crumpling of the interface after a few moments, at $t = 8$ $ms$ (see Fig. 1($d$)($ii$)), owing to the presence of shell of clay particles. These above results confirm self-assembly of solid layer of clay particles around the oil drop enabling us to explore the effect of increasing $\varphi$ on the bursting of hexadecane (oil) drops. To this end, we varied $\varphi$ from 1 to 6 wt% and recorded the number of bursting events, $N_f$ required to arrest the daughter droplet cascade. The sequence of images obtained from videos (Video 1, Multimedia available online) recorded from our experiments is shown in Fig. 1($e$)($i$)-($iv$). The first row represents the case when particles were absent in the oil drop and it is observed that after $N_i = 5$ the entire original parent oil drop formed a film (without clay particles) as reported in our previous work[17]. On introducing clay particles at increasing $\varphi$ of 1%, 3% and 6%, as shown in the second, third and the fourth row, the cascade is arrested much earlier, at $N_f = 4, 3$ and 2 respectively (Video 1, Multimedia available online) producing a particle covered drop, reminiscent of a *Pickering* drop of the anti-Bancroft type[23,35,36] each time, which was stable for atleast 18 hours until the water bath completely evaporated.

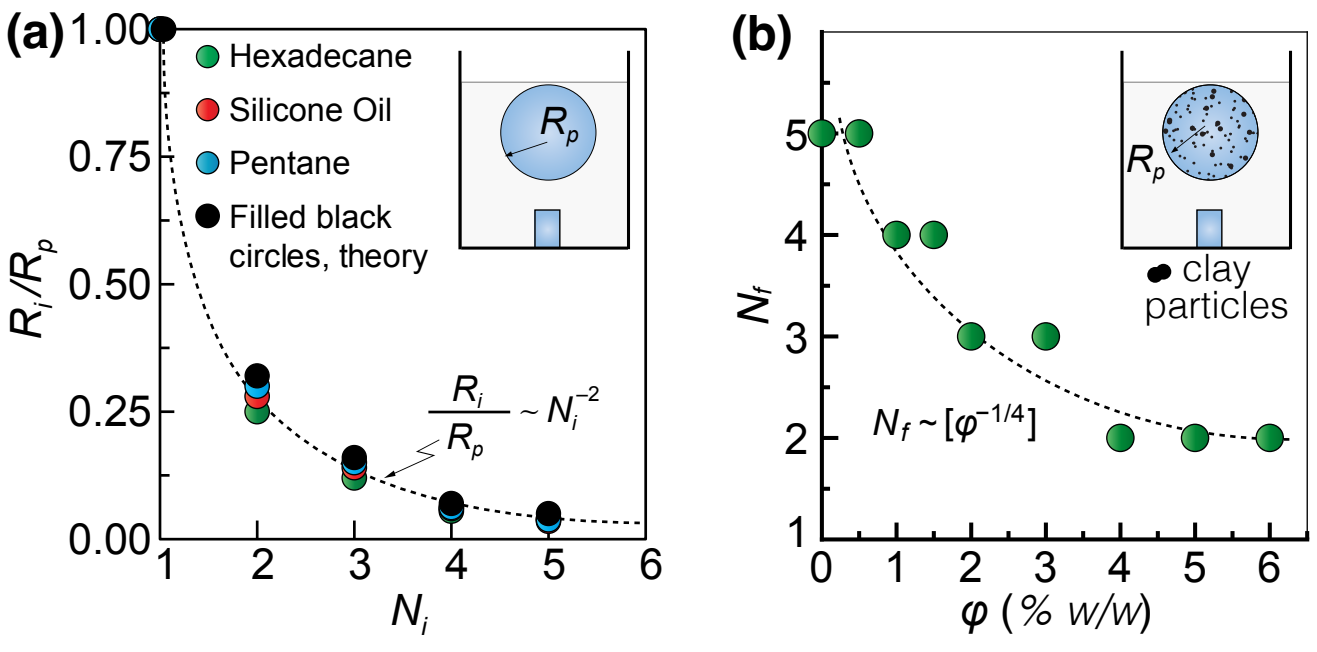

FIG. 2. ($a$) Ratio of daughter droplet $R_i$ and parent drop radius, $R_p$ at each bursting event, $N_i$ in the absence of particles in three oils (hexadecane, silicone oil, pentane)[17] drop yielding the scaling, $R_f/R_p \sim N_i^{-2}$. ($b$) Decrease in number of bursting events required arrest cascade $N_f$ with increasing clay particle concentration ($\varphi$) in the oil drop exhibiting a scaling dependence of the form, $N_f \sim [\varphi^{-1/4}]$. Refer Fig. 1($e$) and ($f$) for symbols used here.

In order to understand the underlying mechanism, we consider the particle-laden parent oil drop of radius $R_p$ after it is released from the nozzle as portrayed schematically in Fig. 1($f$)($i$) and denoted by the symbol, $N_i = 1$, where the subscript, $i$ denotes the generation of daughter droplet with $i = 1$ being the parent drop.

During the rise of the parent oil drop, clay particles preferentially self-assemble[15,35] along the water-oil interface with a certain initial surface coverage, $\chi_1$ ultimately halting near the water-air interface. This is followed by bursting of thin bulk ($b$) water film between the oil drop and water-air interface producing a daughter droplet[17] (of radius, $R_i$) with higher intermediate particle surface coverage, $\chi_i$ compared to the initial particle-laden parent drop and a thin layer particle-infused oil film atop the water-air interface as sketched in Fig. 1($f$)($ii$). With subsequent bursting events ($N_i > 1$) the daughter droplet size, $R_i$ continues to decrease with increasing $\chi_i$ eventually resulting in tight enough packing of particles (with coverage $\chi_f$) on its drop/bulk interface at which stage ($N_f$) subsequent bursting is arrested. No more oil can drain through this particle shell at this stage and bursting is jammed by the particles-infused oil film residing above (see Fig. 1($f$)($iii$)). Even though it might follow from (initial) high value of $\chi_1$ that bursting will not commence, such a configuration does not lead to stable oil-in-water emulsion, restricting $\varphi$ to 0 to 6 wt. % (see SM, Sec. 1 for details and clay particle size distribution). Using the expression by Golemanov[36], $\varphi = 8\chi_1(\rho_c/\rho_o)/((2R_p/\langle a_c\rangle) + 8\chi_1(\rho_c/\rho_o - 1))$ and plugging in values of $\chi_1 = 0.9$, $\rho_c = 2400$ kg·m$^{-3}$, $\rho_o = 773$ kg·m$^{-3}$, $R_p = 2.2$ mm, $\langle a_c\rangle = 15$ $\mu$m for anti-bancroft emulsions we obtain, $\varphi \approx 7\%$ for which a drop of radius 2.2 mm will be stable by itself. We have chosen a limit slightly below this value of 6% to ensure we continue to see atleast one bursting event.

To make quantitative predictions about $N_f = F(\varphi)$ based on the above mechanism, we note that until the formation of a final stable daughter droplet of radius, $R_f$, the bursting process is expected to be similar to the case when particles are absent (see Fig. 1($e$)($i$)) since $\chi_{i=1}$ is not large enough to cover the entire drop's surface initially. However, once $\chi_i$ reaches $\chi_f = 0.90$, the cascade is arrested[36] by the mechanism explained previously. We could use this to relate $R_f$ with final number of particles, $n_{c,f}$ that it contains using the relation[36], $\chi_f = n_{c,f}\langle a_c\rangle^2/16R_f^2$. For a constant $\langle a_c\rangle$ and $\chi_f = 0.9$ (at arrest) we may write, $R_f^2 \sim n_{c,f}$. From here, it only remains to connect $n_{c,f}$ with $\varphi$ and $R_f$ with $N_f$ to get the functional form of $N_f$ in terms of $\varphi$.

In pursuit of the above, we first recognize, $\varphi = \rho_c\langle a_c\rangle^3 n_{c,i=1}/8\rho_o R_p^3$ where, $n_{c,i=1}$ the initial number of clay particles and $\rho_c \approx 2400$ kg·m$^{-3}$ is density of clay particles[37]. Since, $R_p \approx 3.4$ mm, $\langle a_c\rangle$ and $\rho_o$ are constants for our experiments we obtain, $\varphi \sim n_{c,1}$. Moreover, larger $\varphi$ requires larger $R_f$ with a larger surface area, $\sim R_f^2$ (see previous paragraph) to accommodate all the particles, therefore, it is reasonable to expect, $\varphi \sim n_{c,1} \sim n_{c,f} \sim R_f^2$ thereby simplifying our scaling relation to the form, $R_f^2 \sim \varphi$.

Now, it only remains to connect the number of bursting events, $N_f$ required to arrest the cascade at a specific daughter droplet of radius, $R_f$. To do so, we use the scaling relation derived in our previous work[17] for the daughter droplet size, $R_{i=2} \sim \zeta^{-2.34} R_p^{2.74}$ where the constant, $\zeta = \mu_b/\sqrt{\rho_b \sigma_{pb}}$. To obtain the daughter droplet size after the next bursting event, $N_i = 3$ we substitute $R_{i=2}$ from this scaling as $R_p$ in the same expression. Recursively, this gives rise to, $R_{i=3} \sim \zeta^{-2.34(1+2.74)} R_p^{2.74^2}$. Therefore, after $N_i\,(\geq 2)$ cascades we write the general expression for $R_i$, the daughter droplet size after $N_i$ generations, $R_i/R_p \sim \zeta^{-2.34\mathcal{M}} R_p^{2.74^{N_i-1}-1}$, where, $\mathcal{M} = \sum_{i=1}^{N_i-1} 2.74^{i-1}$. We use this expression to determine theoretically the variation of the daughter droplet size, $R_i/R_p$ with $N_i$ for a given $R_p \approx 3.4$ mm which we see to match our experimental values accurately (see Fig. 2($a$)). However, this form is inconvenient to use for further analysis and therefore we seek a more concise functional dependence of the type, $R_i/R_p = F(N_i)$. To do so, we use a best fit curve to represent our experimentally and theoretically obtained values and obtain the relation, $R_i \sim R_p N_i^{-2}$ which for a constant $R_p \approx 3.4$ mm reads, $R_i \sim N_i^{-2}$. This relation could alternatively even be obtained rigorously by expanding $\zeta^{-2.34\mathcal{M}} R_p^{2.74^{N_i-1}-1}$ in Taylor series to obtain its polynomial form but not undertaken here for conciseness. At the arrest radius, $R_f$ this assumes the form, $R_f \sim N_f^{-2}$, obtained by setting $i = f$ in $R_i \sim N_i^{-2}$. Finally, we combine this scaling relation with $R_f^2 \sim \varphi$ which gives number of bursting events or cascades, $N_{i=f}$ before arrest at an initially prescribed $\varphi$ as,

$$N_f \sim [\varphi^{-1/4}] \quad (1)$$

In Eq. 1 [·], is the nearest integer function and the scaling obtained accurately predicts the data as shown in Fig. 2($b$) with a scaling prefactor evaluated as $1.2 \pm 0.03$. We also state that the scaling relation (Eq. 1) remains unchanged for any particle size distribution (see Sec. S1 SM) or any initial parent drop radius, $R_p$ as it algebraically cancels out leading to Eq. (1).

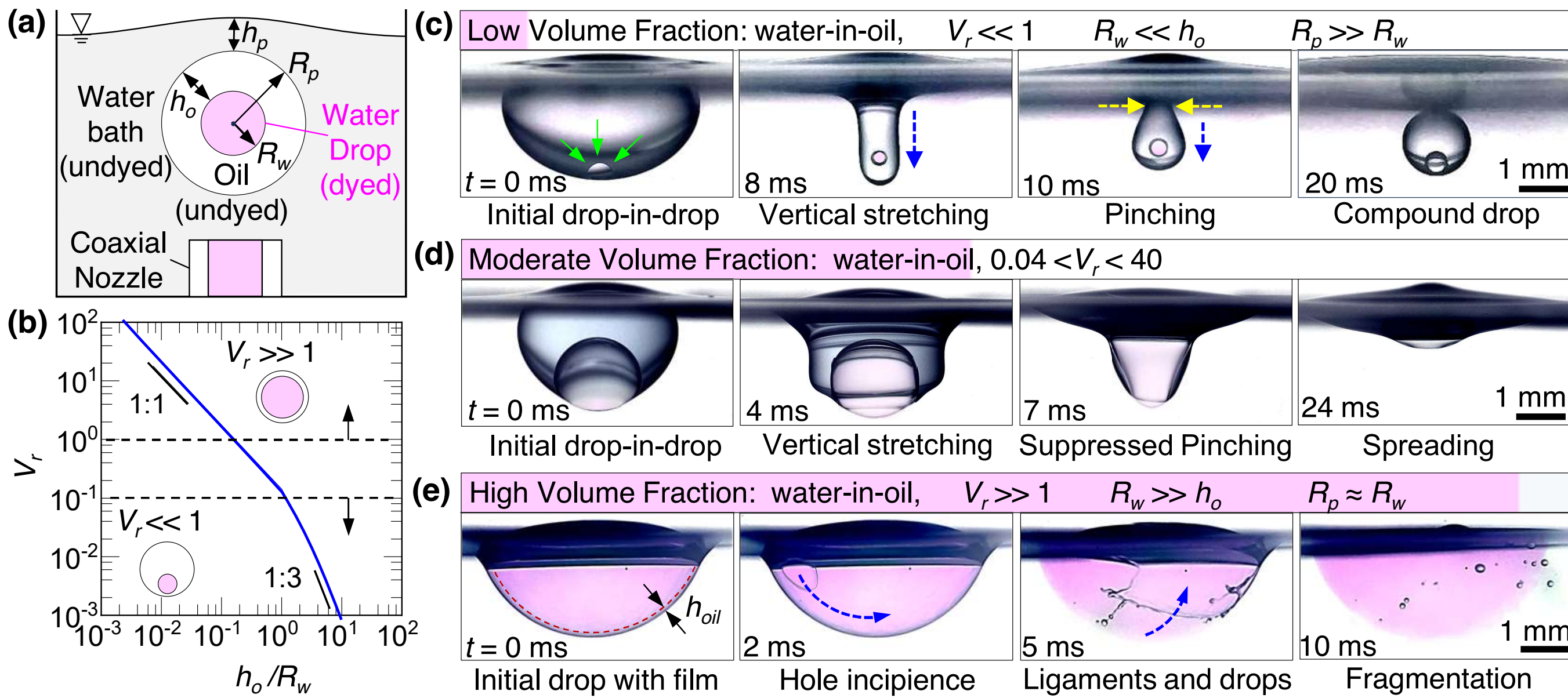

FIG. 3. (*a*) Schematic shows a rising water-in-oil compound drop generated using a co-axial nozzle. The inner water drop is dyed pink using rhodamine D while the water bath and the oil covering it are not. (*b*) Dependence of volume ratio, $V_r$ of water-in-oil in a compound drop on the ratio of oil film thickness and water drop radius, $h_o/R_w$. In the limit, $V_r << 1$ it reduces to $V_r \approx (h_o/R_w)^{1/3}$ and for $V_r >> 1$ it yields, $V_r \approx (3h_o/R_w)^{-1}$. Bursting process of a compound drop (Multimedia available online) (*c*) at low $V_r \approx 5\%$ (green arrows, initial encapsulated water drop) producing an encapsulated daughter water drop of higher $V_r$ (dotted arrows, movement direction) (*d*) at intermediate $0.04 < V_r < 40$ with suppression of daughter droplet production. (*e*) at high $V_r \approx 97\%$, leading to underwater oil film fragmentation (dotted arrows, hole expansion direction) and polydispersed daughter oil droplets. Note, waves after bursting approximately travel a distance of $1.5\pi R_p$ to cover the entire drop.

Our results so far show that solid shell of particles provide an attractive route to modify post-bursting outcomes. Encouraged by these findings, we examine the consequence of bursting at an air-water interface when a liquid shell of variable thickness surrounds around the oil drop as rationalized by a water-in-oil-compound drop. To study this we use a coaxial nozzle of inner diameter 0.56 mm and annular gap of 1.7 mm nozzle producing a compound parent drop of radius, $R_p = 2.2$ mm and volume $V_p$ with varying water ($w$) drop radius $R_w$, (0.5-2.1 mm) and volume, $V_w$ producing an oil ($o$) layer of thickness, $h_o$ and volume $V_o$ as shown schematically in Fig. 3(*a*) which is subjects the bulk (water) and oil layer inside the drop to continual drainage (see Sec. S2 SM). Three volume fractions, $V_r := V_w/V_o$ which correspond to $V_r << 1$, $V_r >> 1$ and $0.04 < V_r < 40$ of the compound drops are tested whose dependence on dimensionless oil layer thickness, $h_o/R_w$ is shown in Fig. 3(*b*). For these calculations we compute, $V_r$ as $R_w^3/(R_p^3 - R_w^3)$. Considering $R_p = R_w + h_o$ we can rewrite $V_r$ as, $[(h_o R_p^2/R_w^3)(1 + R_w/R_p + R_w^2/R_p^2)]^{-1}$. In the limit, $V_r << 1$ and $V_r >> 1$ this reduces to, $h_o/R_w \approx V_r^{-1/3}$ and $h_o/R_w \approx \frac{1}{3}V_r^{-1}$ respectively (more details of this algebra may be found in Sec. S3, SM)

Our experimental results at different $V_r$ are shown in Fig. 3(*c*)-(*e*) (also see Video 2, Multimedia available online). At low, $V_r$ ($<< 1$) once the parent compound drop bursts, capillary waves descend downwards pinching the drop to form a daughter droplet which encapsulates the original water drop within (see Fig. 3(*c*)). In emulsion synthesis where excess material removal has been typically achieved through solvent evaporation[38,39] this could provide a facile solution strategy. At intermediate $0.04 < V_r < 40$ the encapsulated water drop is large enough to suppress daughter droplet formation altogether as shown in Fig. 3(*d*) and drop just floats up eventually. Finally, at large, $V_r$ ($>> 1$) the thin oil film surrounding the water drop bursts at the bulk water-air interface producing sub-surface polydispersed oil droplets (see Fig. 3(*e*)) bearing similarities to bursting of curved thin liquid films[40,41] and those produced by raindrop impacts on oil slicks[42].

To mathematically determine when each of these regimes will be observed we develop a design map based on our experimental data as shown in Fig. 4. Our arguments follow from the premise that rising water-in-oil compound drops experiences two competing drainage flows whose time scales are determined by Stefan-Reynolds theory[43](see Sec. S2, SM). At $V_r << 1$, the time scale of drainage, $\sim \mu_o R_w^4/F_{w/o} h_o^2$ of the oil layer ($h_o >> R_w$) due to apparent weight ( $F_{w/o}$) of the encapsulated water drop within the oil (parent, hexadecane) drop given by, $(\rho_w - \rho_o)V_w g + \rho_w V_p g$ (where, $g = 9.81$ m·s$^{-2}$ is gravitational acceleration) is required to be greater than the time scale, $\sim \mu_w R_p^4/F_{p/b} h_p^2$ of buoyancy driven squeezing of bulk water film of thickness $h_p$ driven by the force, $F_{p/b} = (\rho_w - \rho_o)V_p g$ to ensure water drop remains inside the compound daughter droplet after bursting. This condition leads us to the balance, $(\rho_r - 1)(2\rho_r V_r - V_r + \rho_r)^{-1} \sim \mu_r (h_o^2/R_w^2)(R_w^2/h_p^2)$ at the regime boundary, where, $\rho_r := \rho_w/\rho_o$ and $\mu_r := \mu_w/\mu_o$. Using the geometric approximation, $h_o/R_w \approx V_r^{-1/3}$ (see Fig.3(*b*)) and algebraic simplification, $2\rho_r V_r - V_r + \rho_r \approx \rho_r$ for $V_r << 1$ (see Sec. S3, S4, SM) and considering that the fluid properties, $\rho_r = 1.2$, $\mu_r = 3$ and $R_p = 2.2$ mm) are a constant we arrive at the following sim-plified relation for the boundary of encapsulation,

$$R_w/h_p \sim V_r^{1/3} \tag{2}$$

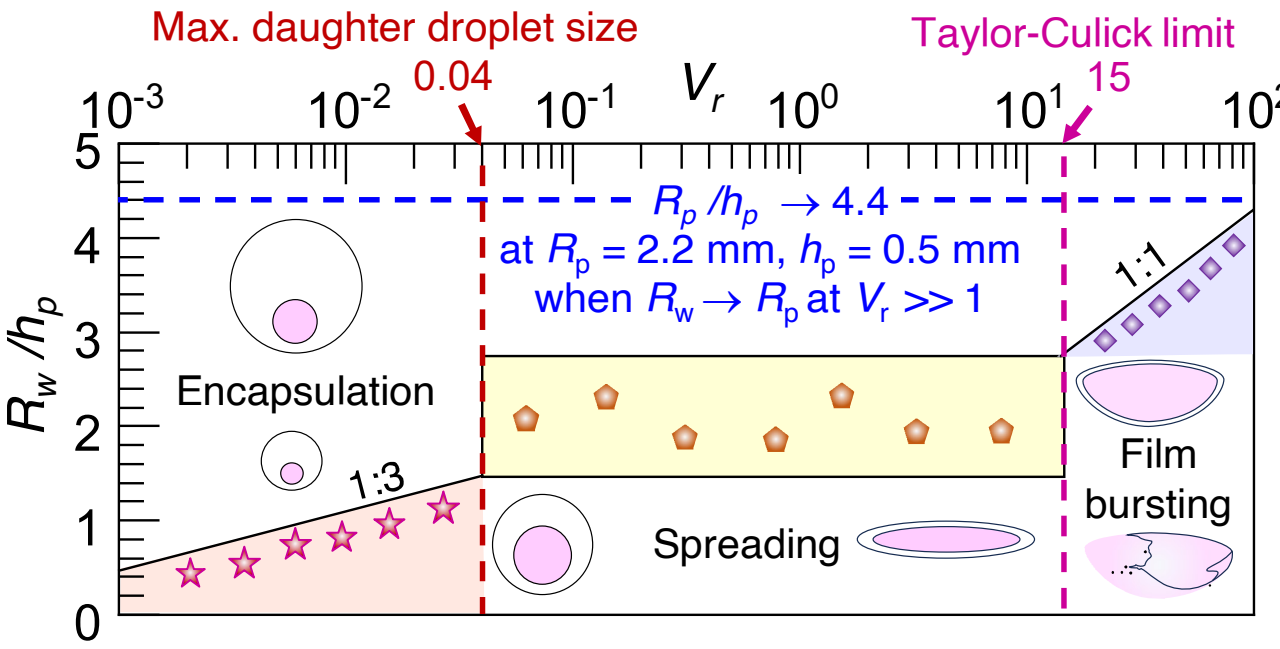

FIG. 4. Design map predicting encapsulation, spreading and film bursting at different dimensionless water drop radius, $R_w/h_p$ as a function of $V_r$ for $R_p = 2.2$ mm. Symbols are experimental data.

Similarly, to derive the criterion of film bursting at $V_r >> 1$ the time scale of drainage of the *thin* oil film ($h_o << R_w$) due to dominant interfacial tension or capillarity (*cap*), $F_{cap} = \sigma_{w/o}\left(2\pi R_p\right)$ should be greater than the buoyancy driven drainage due to force, $F_{p/b}$ described above which at the regime boundary reads, $F_{p/\mathrm{b}}/F_{\mathrm{cap}} \sim \mu_r h_{\mathrm{o}}^2/h_p^2$. This simplifies to, $(\rho_r - 1)(2/3)[(\rho_r - 1)(1+V_r)]^{-1}(Bo_{w/\mathrm{o}}) \sim \mu_r h_{\mathrm{o}}^2/h_p^2$ where, $Bo_{w/\mathrm{o}} = (\rho_w - \rho_o)R_p^2 g/\sigma_{w/o}$. Like above, fluid properties, $\mu_r$, $\rho_r$ and $R_p$ are a constant which along with the geometric approximation, $h_o/R_w \approx (3V_r)^{-1}$ (see Sec. S3, S4, SM and Fig. 3(*b*)) yields at the transition boundary,

$$R_w/h_p \sim V_r \tag{3}$$

For closure, we evaluate three more bounds for $R_p \approx 2.2$ mm: (I) since lubrication flow approximation is valid for Reynolds number, $Re_b = \rho_w v_w h_p/\mu_w << 1$; for a rise velocity, $v_w \approx 0.002$ m/s (from experiments) we obtain $h_p = 0.5$ mm which results in, $R_p/h_p \approx 4.4$. Note that maximum value attained by $R_w/h_p$ is when the encapsulated water drop occupies the entire volume of the compound drop, at $V_r >> 1$, equivalent to $R_w \to R_p$ and therefore, $R_w/h_p = R_p/h_p \approx 4.4$. (II) to ensure complete encapsulation after bursting we require, $R_w \leq R_{i=2} = R_p/2^2$ where, $R_2$ is the daughter droplet radius after first bursting event (see Fig. 2). For, $h_o/R_w \approx V_r^{-1/3}$ at $V_r << 1$ and $h_o/R_w \approx R_p/R_w$ this results in, $V_r \approx 0.04$ (III) Lastly, at $V_r >> 1$ the oil film transitions from capillary dominated dynamics at moderate $V_r$ at a time scale[28], $\tau_{mod} = (\rho_o R_p^3/\sigma_{w/o})^{0.5}$ with capillary waves traveling a distance $\approx 1.5\pi R_p$ at an average velocity scale, $v_{mod}/2 = 1.5\pi R_p/\tau_{mod}$ (see Fig. 3(e)) to bursting of a curved film[44] at higher $V_r$ with an expanding hole retracting at the Taylor-Culick velocity[44], $v_{tc} = (2\sigma_{w/o}/\rho_o h_o)^{0.5}$. For smooth crossover, $v_{tc}$ equals $v_{mod}$ which simplifies to, $h_o/R_p = (9\pi^2/2)^{-1}$. On using the geometric approximation $V_r \approx (3h_o/R_w)^{-1}$ for $V_r >> 1$ and $R_w \approx R_p$ we ultimately obtain the bound, $V_r \approx 15$. The limits (I), (II) and (III) (see Sec. S4, SM for details) along with Eqs 2 and 3 shown in Fig. 4 complete the design map.

In summary, we show two ways to tailor consequence of a bursting oil drop at an air-water interface. First, we introduce hydrophilic clay particles inside the oil drop to arrest daughter droplet generation and, second, we encapsulate a water droplet inside an oil drop to find distinct behaviors at various water to oil volume fractions. The former of our demonstrations is directly connected to practical situations like entrainment of clay particles from the ocean bed in underwater oil pipeline bursts and *Pickering* emulsions. The latter is inspired by water encapsulation in buoyant oil jets, optimizing size of double emulsions and raindrop impact on oil slicks. In practical scenarios as seen in oceans and seas, we expect humidity, temperature, surfactants and other contaminants to be present[45]. These can delay or accelerate the bursting processes but not prevent their observance altogether as reported in our work. Hence, we do expect our results to be relevant in normal marine climatic conditions and serve as guide for any future detailed investigations. In addition to oceanic/atmospheric sciences and colloidal synthesis for drug delivery/food/cosmetics, applications that benefit from ingenious manipulation of the consequences of bursting drops could also find our results useful.

## Acknowledgment

Financial support for this project through NSF (EAGER) award no. 2028571 is gratefully acknowledged. The authors thank Navid Saneie for his assistance in obtaining the SEM image for bentonite clay particles.

## Supplementary Material (SM)

See the supplementary material (SM) accompanying this manuscript that contains (*i*) consequences of high initial particle coverage in arresting daughter droplet formation and effect of particle size distribution (*ii*) additional details of algebra leading to the scaling inequalities, (2) and (3) and bounds, I, II and III.

## Data Availability Statement

Data underlying the conclusions of the paper are available in the plots presented and can be provided upon request.

## Conflict of Interest Statement

The authors have no conflict of interests to disclose.